\documentclass[intlimits,twoside,a4paper]{article}

\usepackage{amsmath,amssymb}
\usepackage{epstopdf}
\usepackage{graphicx}
\usepackage{graphics}
\usepackage[T2A]{fontenc}
\usepackage[cp1251]{inputenc}

\usepackage[eqsecnum]{cmpj3}

\issue{2024}{27}{2}{23801}
\doinumber{10.5488/CMP.27.23801}

\title
{Self-assembly behaviour of diblock copolymer--diblock copolymer under oscillating shear field}

\author{Y. Guo\orcid{0000-0003-1576-5148}\refaddr{label1}, H. He\orcid{0000-0006-0427-2574}\refaddr{label2}, X. Fu\orcid{0000-0009-2530-6915}\refaddr{label2}}

\addresses{
\addr{label1} Department of Chemical Engineering and Materials Engineering, Lyuliang University, Lishi, 033001, China
\addr{label2} School of Chemistry and Chemical Engineering, Shanxi University, Taiyuan, 030006, China
}

\date{Received February 03, 2024, in final form May 14, 2024}

\begin{document}

\maketitle

\begin{abstract}
The self-assembly behaviour of a diblock copolymer--diblock copolymer mixture under an oscillating shear field is investigated via cell dynamics simulation. The results indicate that the macrophase separation of the com\-po\-si\-te system is accompanied by the corresponding microphase separation induced by the oscillating shear field. With an increase in the shear frequency, the AB phase changes from a tilted layered structure to a parallel layered structure, and finally to a vertical layered structure. The CD phase transforms from the initial concentric ring into a parallel layer in the ring and then into a parallel layered structure; thus, the system finally forms a layered structure of the AB phase (vertical layer) and CD phase (parallel layer) perpendicular to each other. To verify the phase transition, the dynamic evolution of the domain size at different shear frequencies is analysed. The ordered phase transition with an increase in the oscillating shear field varies when the initial composition ratio of the system is changed. This conclusion provides a valuable guidance for the formation and transformation of ordered structures in experiments.

%

\end{abstract}

\section{Introduction}

In recent decades, diblock copolymers (DBCs) have attracted considerable attention because of their capability to self-assemble into various amazing structures and their applications in nanotechnology, microelectronics, and clean energy~\cite{key1,key2,key3,key4,key5}. In addition to their novel structures, the formation of the ordered structure and the ordered phase transformation of DBCs are important for the development of new functional materials~\cite{key6,key7,key8,key9,key10}. If the homopolymer~\cite{key11,key12,key13,key14} or a different DBC~\cite{key15,key16,key17} is added, a multiscale hierarchical structure can be formed, and its function can be optimised while enriching the structure of the system.

Many theoretical studies have been performed on the doping of a DBC with a homopolymer or a different DBC. Both phenomenological methods~\cite{key18} and numerical simulations~\cite{key19} have yielded results consistent with experiments, while predicting new structures and properties. Martinez et~al. studied the stabilisation of multiple ordered bicontinuous phases in blends of a DBC and a homopolymer via a combination of particle-based simulations and self-consistent field theory (SCFT)~\cite{key20}. Liu et~al. adopted Monte Carlo simulations to study the influences of the chain length and DBC concentration on the interfacial properties between two immiscible homopolymers~\cite{key21}. Xie et~al. systematically studied the formation and relative stability of spherical packing phases in binary blends composed of AB-DBCs and A-homopolymers by applying SCFT to the freely-jointed chain model of polymers~\cite{key22}. In addition, many researchers have experimentally studied block copolymers doped with homopolymers. Habersberger et~al. systematically described the phase behaviour of a variety of symmetric CE/C/E ternary copolymer/homopolymer blends, where C represents poly(cyclohexylethylene) and E represents poly(ethylene), verifying the location in the composition of the technologically important bicontinuous microemulsion (B\textmu{}E) channel as a function of molecular weight of the diblock~\cite{key11}. Later, Robert et~al. from the same research group discussed a connection between structure and ionic conductivity in salt-containing ternary polymer blends consisting of a polystyrene-poly(ethylene oxide) (PS-PEO) DBC and a poly(ethylene oxide) (PEO) hompolymer~\cite{key12}. Zheng et~al. systematically investigated the phase behaviour of partially charged DBC--homopolymer ternary blends using small-angle X-ray scattering~\cite{key13}. For the system comprising two DBCs, Wang et~al. assessed a versatile computational strategy involving cooperative assembly of DBC blends to form spherical and cylindrical compartmentalised micelles with complicated morphologies and structures. The co-assembly tactic combines the advantages of polymer blending and incompatibility-induced phase separation~\cite{key15}. Su \emph{et~al.} explored the self-assembled structure of a poly(styrene-b-vinylbenzyl triazolylmethyl methyladenine) (PS-b-PVBA) DBC and a poly(vinylbenzyl triazolylmethyl methylthymine) (PVBT) homopolymer via nuclear magnetic resonance spectra, small-angle X-ray scattering and transmission electron microscopy~\cite{key23}. Fan et~al. investigated the phase behaviour of an AB/CD block copolymer blend via SCFT. The results indicated that the phase transitions from layered structures on different spatial scales to a core--shell structure were ascribed to an increase in the fusion degree of components B and D~\cite{key24}. Sun et~al. explored the orientational order transition of striped patterns in microphase structures of DBC--DBC mixtures in the existence of periodic oscillatory particles~\cite{key16}. By altering the oscillatory frequency and amplitude, the orientational order transition of a striped microphase structure from the state parallel to the oscillatory direction to that perpendicular to the oscillatory direction was obtained. However, Pan et~al. made the two DBCs under parallel wall confinement and studied the effects of the distance between walls, the wall--block interaction, and the repulsive interactions between different monomers on the phase behaviour~\cite{key25,key26}.

It is important to understand how to regulate the orientation of ordered structures and how to transform the orientation order more simply, conveniently, and quickly to obtain various self-assembled structures of composite materials. The principal methods of regulating nanostructures are substrate induction, space confinement, and external field application \cite{key27,key28,key29,key30,key31}, all of which can render the bulk more novel and organised. Many studies have proven that the application of an external field is an effective method for regulating polymer nanocomposites~\cite{key32,key33,key34}. An oscillating shear field can enable the system to produce various novel and ordered structures and realise transformation and effective regulation of the structures~\cite{key35,key36,key37}. Experimentally determining the formation of certain structures is often tedious; thus, we need computer simulations to provide a more effective guidance for experiments. In this study, the self-assembly behaviour of two different block copolymers under an oscillating shear field was analysed via computer simulations, and an effective manner in forming and changing the orderly structure of the regulation system was obtained. The remainder of this paper is organised as follows: section~2 describes the model and simulation method; section~3 presents the numerical results and discussions; finally, section~4 presents conclusions.

\section{Models and simulation methods}

We consider a phase separating system consisting of two different DBCs, which are under the oscillating  shear field. The polymer chain of one DBC consists of A and B monomers, which has a short-range repulsive interaction between them. The polymer chain of the other DBC consists of C and D monomers, which also has a short-range repulsive interaction. The B and D monomers in two different DBCs are mutually exclusive with each other. Furthermore, the hydrodynamic effects dominate in the very final stage of microphase separation in polymer mixtures, so they are neglected in the present model.

For DBC-DBC system, several parameters are defined. We consider that the DBCs are symmetric. Thus, the polymerization degree of A block is equal to that of B block, which is also true for the C and D blocks, that is, $N_A=N_B$, $N_C=N_D$. In the process of phase separations, fluctuations are predominant, so we should investigate the local volume fractions of monomers A, B, C and D. They are denoted, respectively, by $\phi_A(x,y)$, $\phi_B(x,y)$, $\phi_C(x,y)$ and $\phi_D(x,y)$. The total density $\phi_{A}(x,y)+\phi_{B}(x,y)+\phi_{C}(x,y)+\phi_{D}(x,y)$ is constant under the incompressibility condition. Then, we take $\psi(x,y)=\phi_{A}(x,y)+\phi_{B}(x,y)$, $\phi(x,y)=\phi_{A}(x,y)-\phi_{B}(x,y)$, and $\xi(x,y)=\phi_{C}(x,y)-\phi_{D}(x,y)$ are the independent variables that are used to characterize the structure ordering. The order parameter $\phi(x,y)$  describes the local concentration difference between the A and B monomers, the order parameter $\xi(x,y)$ gives the local concentration difference between the C and D monomers.

Here, we use a three-order-parameter model~\cite{key18,key19,key38,key39,key40,key41}, whose free-energy functional of the system is given by
\begin{equation}
    F=F_{L}+F_{S},\label{1}
\end{equation}
the long-range part $F_{L}$ is given by
\begin{equation}
    F_L=\frac{\alpha}{2} \iint {\rm d}{\textbf{\emph r}}\,{\rm d}{\textbf{\emph r}'}G({\textbf{\emph r}},{\textbf{\emph r}' })[\phi({\textbf{\emph r}})-\phi_0][\phi({\textbf{\emph r}'})-\phi_0]+\frac{\beta}{2} \iint {\rm d}{\textbf{\emph r}}\,{\rm d}{\textbf{\emph r}'}G({\textbf{\emph r}},{\textbf{\emph r}' })[\xi({\textbf{\emph r}})-\xi_0][\xi({\textbf{\emph r}'})-\xi_0] ,\label{2}
\end{equation}
where $\alpha$, $\beta$ are all positive constants, refer to the long-range interaction. $G({\textbf{\emph r}},{\textbf{\emph r}'})$ is the Green's function defined by the equation $-\nabla^{2}G({\textbf{\emph r}}, {\textbf{\emph r}'})=\delta({\textbf{\emph r}}-{\textbf{\emph r}'})$, while $\phi_0$ and $\xi_0$ is the spatial averages of $\phi$ and $\xi$. As mentioned before, the two DBCs are symmetric, so $\phi_{0}=0$, $\xi_{0}=0$. The short-range part is more complex than the long-range part, which is given by
\begin{equation}
    F_{S}=\iint{\rm d} x\, {\rm d} y \bigg[\frac{d_1}{2}(\nabla\psi)^2+\frac{d_2}{2}(\nabla\phi)^2+\frac{d_3}{2}(\nabla\xi)^2
    +f_{1}(\psi,\phi,\xi)\bigg],\label{3}
\end{equation}
where the $d_1$, $d_2$ and $d_3$ terms correspond to the surface tensions. The local interaction term $f_{1}(\psi,\phi,\xi)$ could be replaced by $f_{1}(\eta,\phi,\xi)$~\cite{key16,key25,key26,key40,key41}, where $\eta=\psi-\psi_{C}$ with $\psi_{C}$ being the volume fraction of two different DBCs at the critical point of the macrophase separation.

It is obvious that the important physical results will be mainly included in the local interaction $f_{1}(\psi,\phi,\xi)$. For further treatment, we can take its form in a phenomenological approach~\cite{key40,key41,key42} as
\begin{equation}
f_1(\eta,\phi,\xi)=\nu_1 (\eta)+\nu_2 (\phi)+\nu_3(\xi)+b_1\eta\phi-b_{11}\eta\xi
-\frac{1}{2}b_2\eta(\phi)^2+\frac{1}{2}b_{22}\eta\phi(\xi)^2.\label{4}
\end{equation}
In the symmetric case, $b_1=(-\chi_{AC}-\chi_{AD}+\chi_{BC}+\chi_{BD})/4$, $b_{11}=(-\chi_{AC}+\chi_{AC}-\chi_{BC}+\chi_{BD})/4$. In the model, the repulsion between the B monomer and the D monomer is set to be the largest, in others words, $\chi_{BD}>\chi_{BC}$, $\chi_{BD}>\chi_{AD}$, $\chi_{BD}>\chi_{AC}$, while the value of $\chi_{BC}$ is close to the value of $\chi_{AC}$ and only a little larger than $\chi_{AC}$, and $\chi_{AD}\gg \chi_{BC}$. In general, $b_1$, $b_{11}$, $b_2$ and $b_{22}$ are all positive constants, thereinto, $b_1$, $b_{11}$ represent the short-range attraction between polymer monomers, $b_1$ mainly arises from the repulsive interaction between the CD DBC and the B monomers, whereas the $b_2$ and $b_{22}$ originate from the conformational entropy between AB DBC and CD DBC. These two terms decide that the observation of a microphase separation could take place in the DBCs. In fact, $b_2$ indicates that a large absolute value of $\phi(x,y)$ is favorable in the region $\eta(x,y)>0$, while $b_{22}$ implies that a large absolute value of $\xi(x,y)$ is more favorable in the region $\eta(x,y)<0$. Equation~\eqref{4} describes the minimum model of the short-range part of the free energy in DBC-DBC system. In the free energy function, the competing action leads to the phase separation of two DBC blends.

According to the three-parameter model, the dynamic equation of phase separation can be described as the time dependent Ginzburg--Landau (TDGL) equation of coupling the diffusion field and velocity field:
\begin{equation}
      \frac{\partial\eta}{\partial t}+{\textbf{\emph v}}\cdot\nabla\eta =M_\eta\,\nabla^2\frac{\delta F}{\delta\eta},\label{5}
\end{equation}
\begin{equation}
    \frac{\partial\phi}{\partial t}+{\textbf{\emph v}}\cdot\nabla\phi =M_\phi\,\nabla^2\frac{\delta F}{\delta\phi},\label{6}
\end{equation}
\begin{equation}
    \frac{\partial\xi}{\partial t}+{\textbf{\emph v}}\cdot\nabla\xi =M_\xi\,\nabla^2\frac{\delta F}{\delta\xi}, \label{7}
\end{equation}
where $M_\eta$, $M_\phi$ and $M_\xi$ are transport coefficients, $\textbf{\emph v}$ is an external velocity field describing the shear flow profile.

For convenience, we can express the external velocity profile as:
\begin{equation}
  \textbf{\emph v}=(\gamma\omega y \cos(\omega t),0),\label{8}
\end{equation}
where $\gamma$ is the shear amplitude, $\omega$ is the shear frequency. $x$ denotes the oscillating shear flow direction, and $y$ denotes the shear gradient direction.

The numerical solutions of the above model system can be carried out in an $L_x \times L_y$ ($128\times128$) two-dimensional square lattice, by using the cell dynamics simulation approach proposed by Oono and Puri~\cite{key43,key44,key45}. The order parameters for each cell are $\eta({\textbf{\emph n} },t)$, $\psi({\textbf{\emph n}},t)$, where ${\textbf{\emph n}}=(n_x,n_y)$ is the lattice position and $n_x$, $n_y$ are integers between 1 and $L$.
Laplace operator is approximated in the cell dynamics simulation as:
\begin{eqnarray}
\nabla^2\phi(n)=\langle\langle\phi(n)\rangle\rangle-\phi(n),\label{9}
\end{eqnarray}
where $\langle\langle\phi(n)\rangle\rangle$ represents the nearest-neighbor $(n.)$, the next-nearest neighbor $(n..)$ cells of $\phi({n})$~\cite{key46}:
\begin{eqnarray}
\langle\langle\phi(n)\rangle\rangle=\frac{1}{6}\sum_{n=n.}\phi(r)+\frac{1}{12}\sum_{n=n..}\phi(r).\label{10}
\end{eqnarray}
In the lattice size ($\Delta x$, $\Delta y$) and the time step ($\Delta t$) are all set to unity.

The cell dynamics simulation equations corresponding to equations~\eqref{5}--\eqref{7},  in their space-time discreteness form, are written as follows:
\begin{eqnarray}
  \eta(\textbf{\emph r},t+\Delta{t})&=&\eta(\textbf{\emph r},t)-\frac{1}{2}\gamma \sin(\omega t)[\eta(x+1,y,t)-\eta(x-1,y,t)]\nonumber \\
  &&+M_\eta(\langle\langle I_\eta\rangle\rangle-I_\eta),\label{11}
\end{eqnarray}
\begin{eqnarray}
  \phi(\textbf{\emph r},t+\Delta{t})&=&\phi(\textbf{\emph r},t)-\frac{1}{2}\gamma \sin(\omega t)[\phi(x+1,y,t)-\phi(x-1,y,t)]\nonumber \\
  &&+M_\phi(\langle\langle I_\phi \rangle\rangle-I_\phi)-\alpha\phi(\textbf{\emph r},t),\label{12}
\end{eqnarray}
\begin{eqnarray}
 \xi(\textbf{\emph r},t+\Delta{t})&=&\xi(\textbf{\emph r},t)-\frac{1}{2}\gamma \sin(\omega t)[\xi(x+1,y,t)-\xi(x-1,y,t)]\nonumber \\
  &&+M_\xi(\langle\langle I_\xi \rangle\rangle-I_\xi)-\beta\xi(\textbf{\emph r},t),\label{13}
\end{eqnarray}
where
\begin{equation}
    I_\eta=-d_1(\langle\langle{\eta}\rangle\rangle-\eta)-A_\eta\tanh\eta+\eta+b_1\phi-b_{11}\xi-
    \frac{1}{2}b_2\phi^2+\frac{1}{2}b_{22}\xi^2,\label{14}
\end{equation}
\begin{equation}
    I_\phi=-d_2(\langle\langle{\phi}\rangle\rangle-\phi)-A_\phi\tanh\phi+\phi+b_1\eta-b_2\eta\phi,\label{15}
\end{equation}
\begin{equation}
    I_\xi=-d_3(\langle\langle{\xi}\rangle\rangle-\xi)-A_\xi\tanh\xi+\xi-b_1\eta-b_{22}\eta\xi.\label{16}
\end{equation}
A shear periodic boundary condition~\cite{key42,key47,key48,key49,key50,key51,key52,key53,key54} should be applied to $x$ direction,
\begin{equation}
    \phi(n_x,n_y,t)=\phi[n_x+N_{x}L+\gamma(t)N_{y}L,n_y+N_{y}L],\label{17}
\end{equation}
where $N_x$ and $N_y$ are arbitrary integers, $\gamma_{0}(t)$ is the shear strain, and $\gamma_{0}(t)=\gamma \sin(\omega t)$. The parameters are chosen to be $d_1=1.0$, $d_2=0.5$, $d_3=0.5$, $b_1=0.008$, $b_2=0.2$, $b_{11}=0.1$, $b_{22}=0.2$ and $M_\eta=M_\phi=M_\xi=1$, the initial distribution of $\phi$, $\eta$ and $\xi$ are specified by random uniform distributions in the range $[-0.01,0.01]$, according to the previous work. In the present simulation work, in the case of $\alpha=0.04$, $\beta=0.03$, by the formula $\alpha=12/[{N^2}{f_b}(1-f_b)]$ ($f_b$ denotes a block ratio)~\cite{key55}, it can be obtained that $N_A=N_B=17$, $N_C=N_D=20$, where $N_{AB}=N_A+N_B$ and $N_{CD}=N_C+N_D$. In this paper, all parameters are scaled, so all of them are dimensionless.

\begin{figure}[!h]
\centerline{\includegraphics[width=0.85\textwidth]{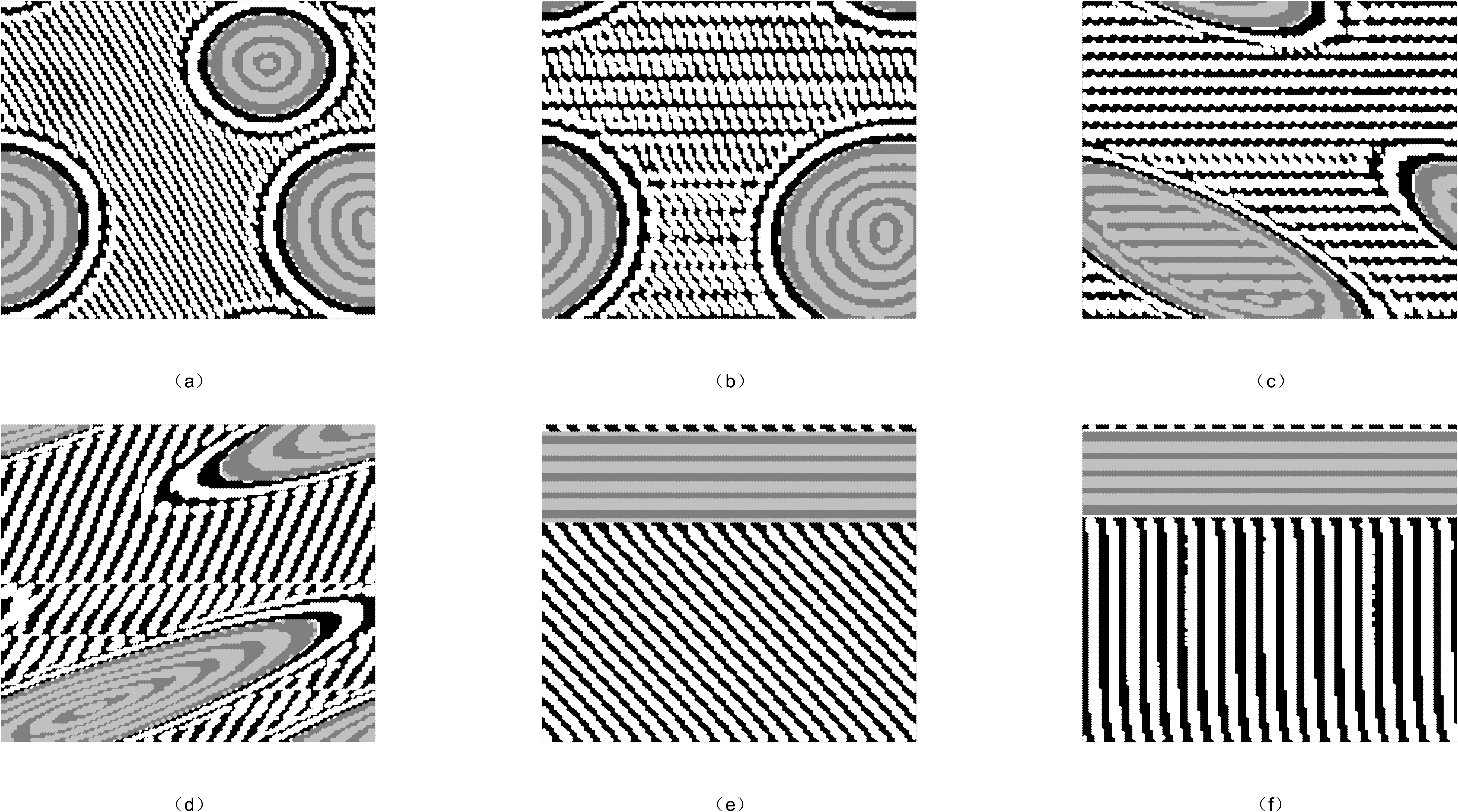}}
\caption{Pattern evolution of a $128\times128$ two-dimensional lattice for $f_{AB/CD}=7/3$, $t=3000000$, $\gamma=0.02$: (a) $\omega=0.0000001$; (b)~$\omega=0.00001$; (c)~$\omega=0.00002$; (d)~$\omega=0.00004$; (e)~$\omega=0.48$; (f)~$\omega=0.66$. White represents phase A, black represents phase B, light gray represents phase C, and dark gray represents phase D.} \label{fig-smp1}
\end{figure}

\section{Numerical results and discussion}
\subsection{Microphase transformation induced by oscillating shear field}

\begin{figure}[!h]
\centerline{\includegraphics[width=0.75\textwidth]{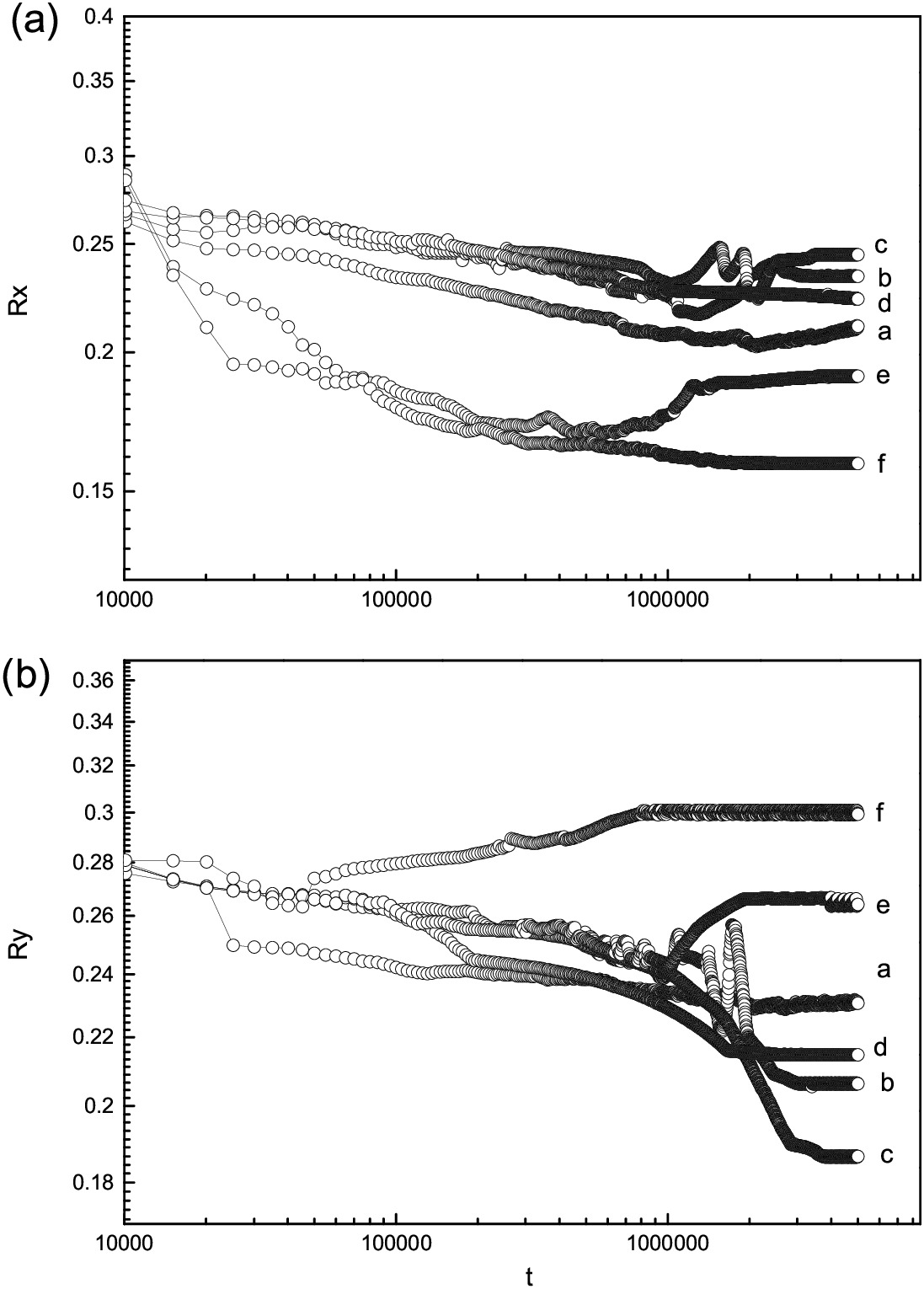}}
\caption{Double-log plots of the domain sizes of the AB DBC at different shear frequencies. (a) Domain size along the $x$-axis $R_x(t)$; (b) domain size along the $y$-axis $R_y(t)$. Curve a, $\omega=0.0000001$; curve b, $\omega=0.00001$; curve c, $\omega=0.00002$; curve d, $\omega=0.00004$; curve e, $\omega=0.48$; curve f, $\omega=0.66$.} \label{fig-smp2}
\end{figure}

First, we study the microphase transformation of DBC--DBC with changes in the shear frequency at $f_{AB/CD}=7/3$, where $f$ represents the ratio of the volume fraction of the AB copolymer and CD copolymer, and $\gamma=0.02$, as shown in figure~\ref{fig-smp1}. In the diagram, the white, black, light gray, and dark gray regions represent phases A, B, C, and D, respectively. For clarity, $\phi>0$ means the A-rich domain, $\phi<0$ means the B-rich domain, $\xi>0$ means the C-rich domain, and $\xi<0$ means the D-rich domain. 

\begin{figure}[!h]
\centerline{\includegraphics[width=0.75\textwidth]{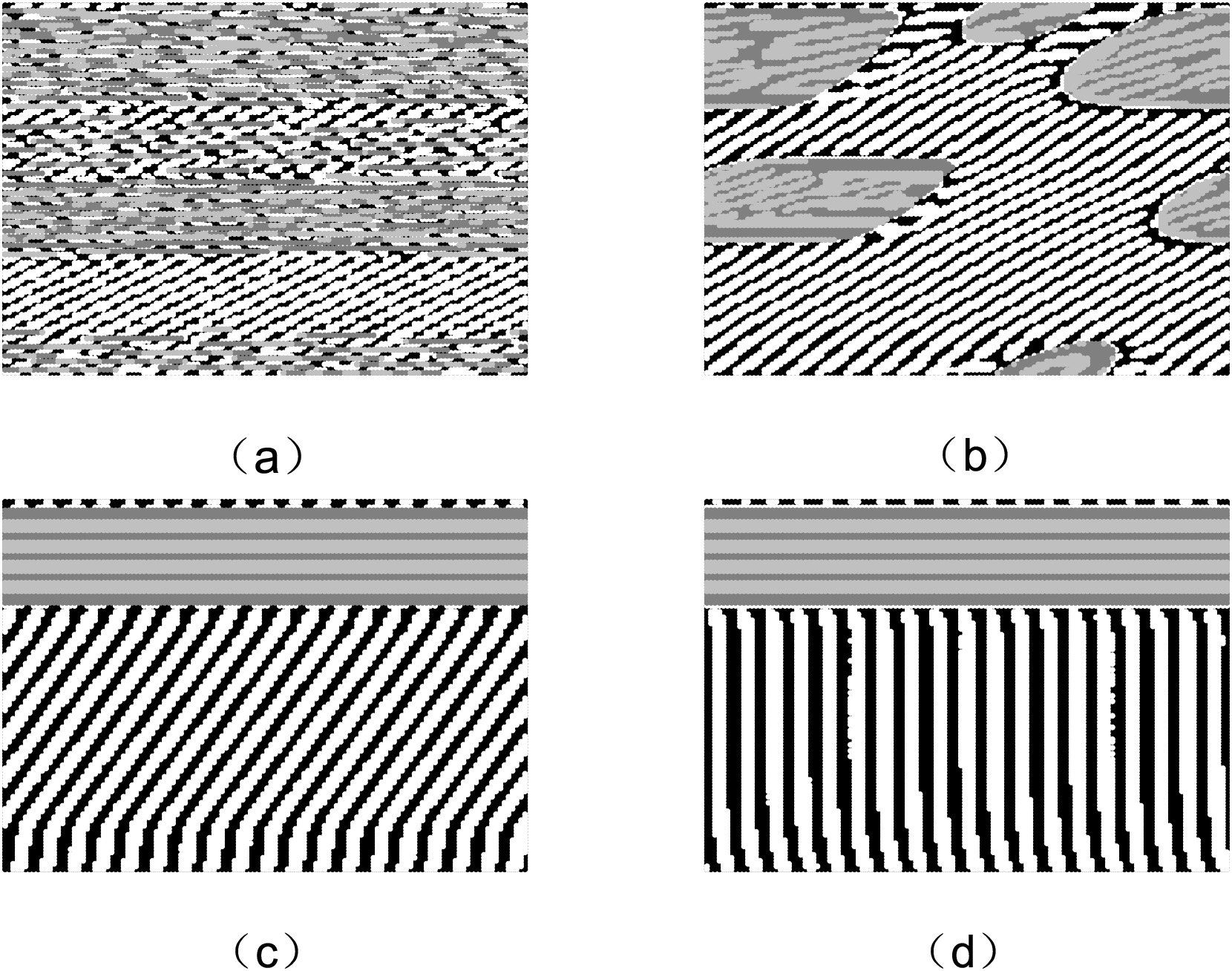}}
\caption{Morphology evolution of the polymer system with $f_{AB/CD}=7/3$, $\gamma=0.02$, $\omega=0.66$, and $b_{22}=0.2$: (a) $t=250000$; (b) $t=500000$; (c) $t=2000000$; (d) $t=3000000$.} \label{fig-smp3}
\end{figure}

In figure~\ref{fig-smp1}, the CD phase is a concentric ring structure, and the AB phase is a tilted layered structure, except for a small amount of AB phase wrapped in the CD phase, which is a concentric ring when the shear frequency is 0.0000001. In other words, the effect of the oscillating shear field on the domain structure of the system is negligible when the shear frequency is low. The result obtained in this case is roughly the same as the phase morphology without an oscillating shear field [figure~\ref{fig-smp1}(a)]. The oscillatory shear begins to take effect as the shear frequency increases to 0.00001. The microphase separation of the system is evidently disturbed, the coarsening degree of the AB phase along the $x$ direction is significantly increased, and the structure of the CD phase is still a concentric ring [figure~\ref{fig-smp1}(b)]. When the shear frequency is increased to 0.00002, the AB phase is almost parallel to the direction of the oscillatory shear, exhibiting a parallel layered structure, while the CD phase transforms from the original concentric ring structure to a parallel layered structure in the ring along the oscillatory shear direction [figure~\ref{fig-smp1}(c)]. When the shear frequency increases to 0.00004, the AB phase transforms back into the tilted layered structure, and the CD phase becomes an inclined concentric ring structure stretched along the oscillatory shear direction [figure~\ref{fig-smp1}(d)]. The CD phase is a parallel layered structure along the $x$ direction, and the AB phase is an oblique layered ordered structure as shear frequency significantly increases to 0.48 [figure~\ref{fig-smp1}(e)]. As the shear frequency increases to 0.66, the CD phase remains parallel lamellae, while the AB phase completely transforms into a vertical layered structure perpendicular to the oscillatory shear direction. Finally, a lamellar structure comprising the AB phase (perpendicular lamellae) and CD phase (parallel lamellae) perpendicular to each other is formed, as shown in figure~\ref{fig-smp1}(f). In summary, as the oscillatory shear frequency increases, the CD phase changes from a concentric ring structure to a parallel layered structure, and the AB phase changes from a tilted layered structure to a parallel layered structure and then to a perpendicular layered structure.

\begin{figure}[!h]
\centerline{\includegraphics[width=0.75\textwidth]{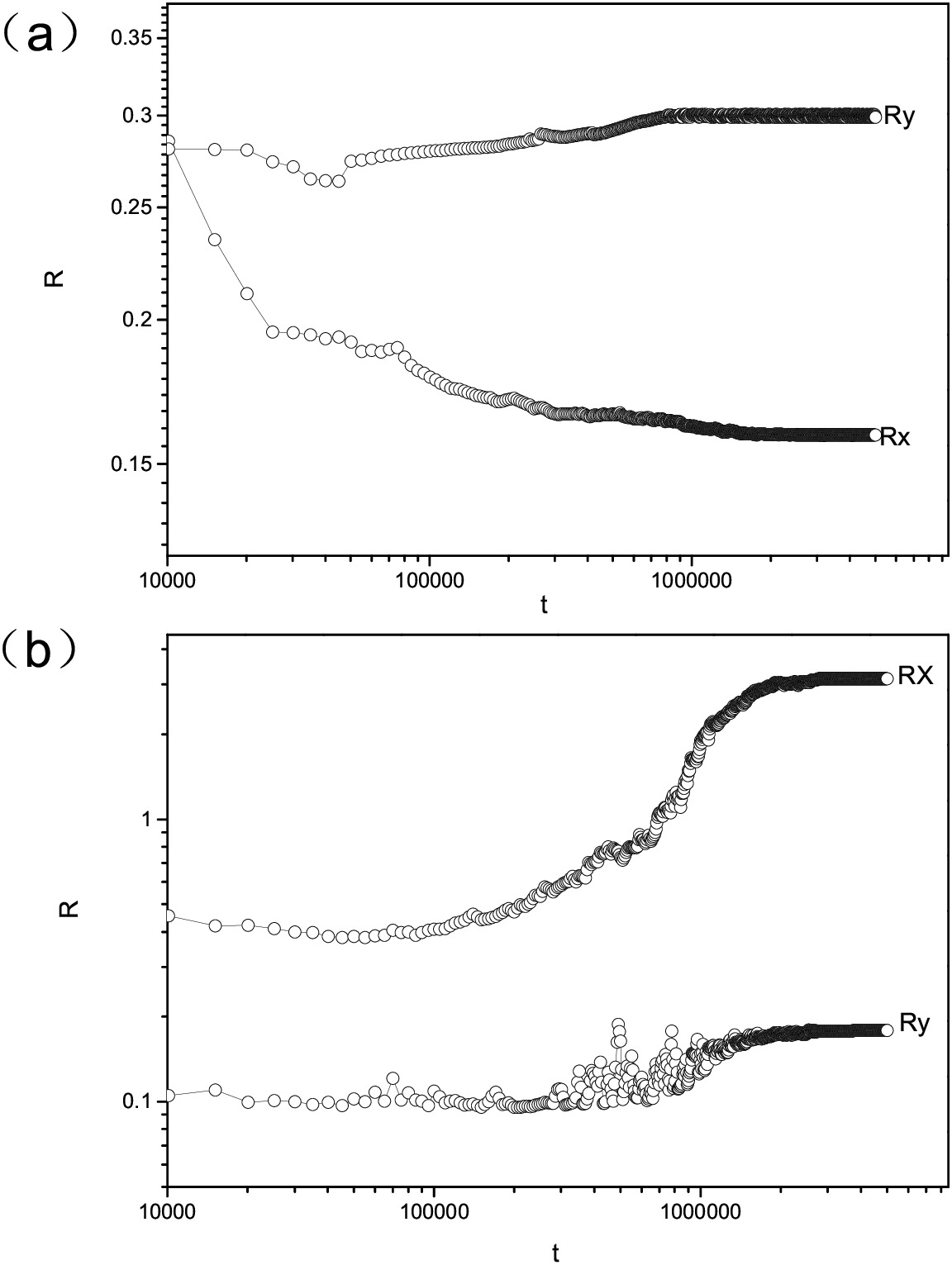}}
\caption{The growth of domain size $R_i(t)$ ($i=x$ or $y$) of the (a) AB phase and (b) CD phase in the $x$ and $y$ directions as a function of time in double-logarithmic plots with $f_{AB/CD}=7/3, \gamma=0.02,\omega=0.66, b_{22}=0.2$.
} \label{fig-smp4}
\end{figure}

These results can be explained as follows. The influence of the oscillating field on the domain structure of the system is negligible when the shear frequency is low. Increasing the shear frequency enhances the coarsening degree of the system in the $x$ direction; therefore, both the AB phase and the CD phase in the ring are almost parallel layered structures along the oscillating field at the appropriate frequency of 0.00002. As the shear frequency continues to increase, the different components push each other and are coarsened in the $y$ direction under the action of rapid periodic oscillatory shear. Under the condition of $f_{AB/CD}=7/3$, the AB phase occupying a larger proportion is significantly affected by the oscillating shear field, which results in an easy pileup in the direction perpendicular to the oscillating shear field, forming the perpendicular layered structure. Since the CD phase is confined in the ring before, it breaks the restraint of the ring at a higher shear frequency and forms a parallel layered structure along the oscillating field.

\begin{figure}[!h]
\centerline{\includegraphics[width=0.75\textwidth]{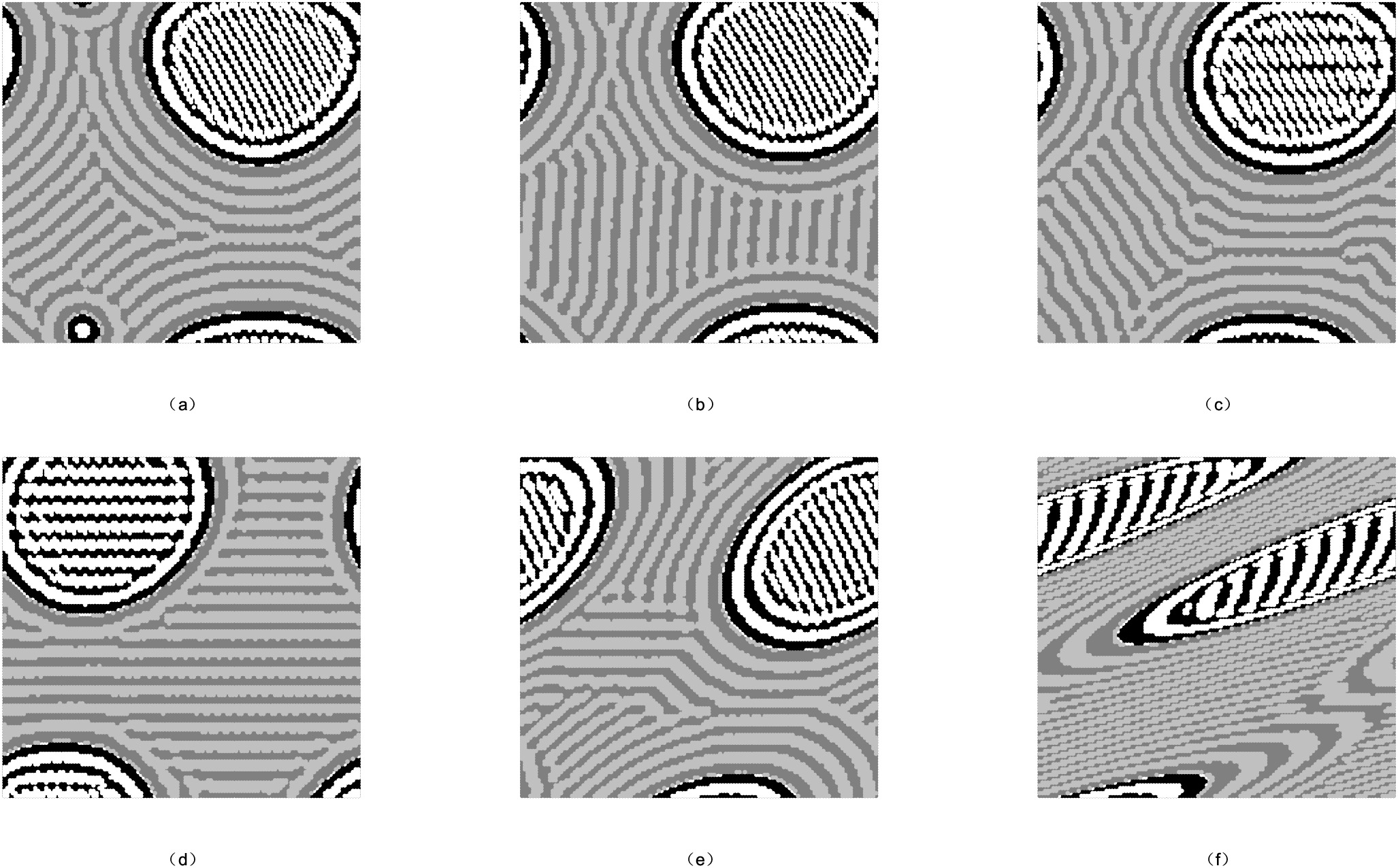}}
\caption{Pattern evolution of a $128\times128$ two-dimensional lattice for $f_{AB/CD}=35/65$, $\omega=0.00001$, and $b_{22}=0.2$: (a) $\gamma=0.00002$; (b) $\gamma=0.00006$; (c) $\gamma=0.022$; (d) $\gamma=0.032$; (e) $\gamma=0.1$; (f) $\gamma=0.6$.} \label{fig-smp5}
\end{figure}

To verify the above microphase transition, we calculate the domain size $R_i(t)$ ($i=x$ or $y$) in the $x$ or $y$ direction as a function of time. The domain sizes $R_i(t)$ can be derived from the inverse of the first moment of the structure factor $S(\textbf{k},t)$ as
\begin{equation}
    R_i(t)=2\piup/\langle k_i(t)\rangle,\label{18}
\end{equation}
where
\begin{equation}
    \langle|k_i(t)|\rangle=\int{\rm d} {\textbf{k}}\,k_iS(\textbf{k},t)/\int{\rm d}{\textbf{k}}\,S({\textbf{k}},t).\label{19}
\end{equation}
The structure factor $S(\textbf{k},t)$ is decided by the Fourier component of the spatial concentration distribution. Figure~\ref{fig-smp2} shows the time evolution of the microdomain sizes $R_i(t)$ in the $x$ and $y$ directions in double-logarithmic plots. The results are the average values for 10 independent runs.

\begin{figure}[!h]
\centerline{\includegraphics[width=0.65\textwidth]{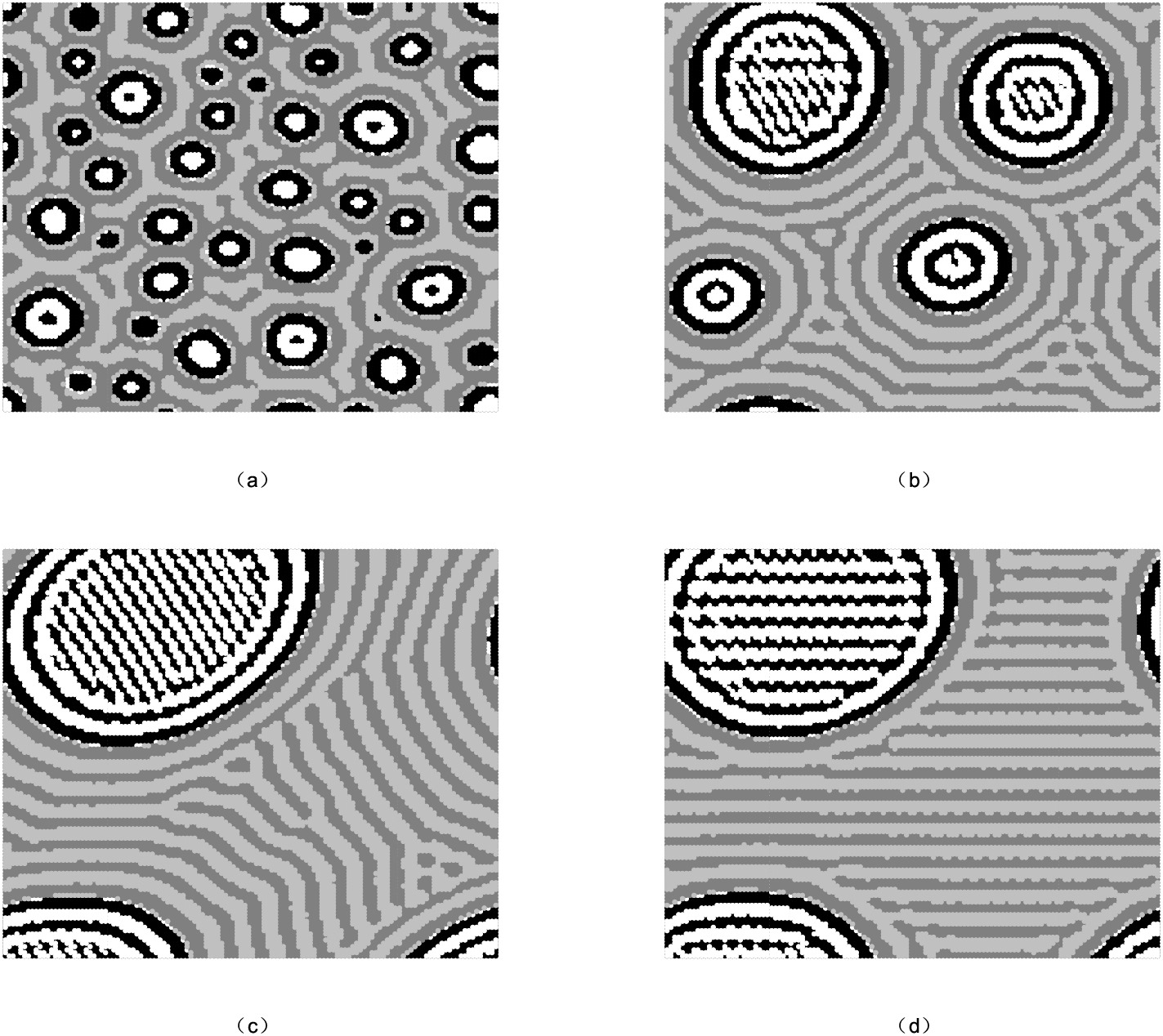}}
\caption{Morphology evolution of the composite system at different times
with $f_{AB/CD}=35/65$, $\omega=0.00001$, and $\gamma=0.032$: (a) $t=10000$; (b) $t=500000$; (c) $t=2000000$; (d) $t=3000000$.
} \label{fig-smp6}
\end{figure}

\begin{figure}[!h]
\centerline{\includegraphics[width=0.65\textwidth]{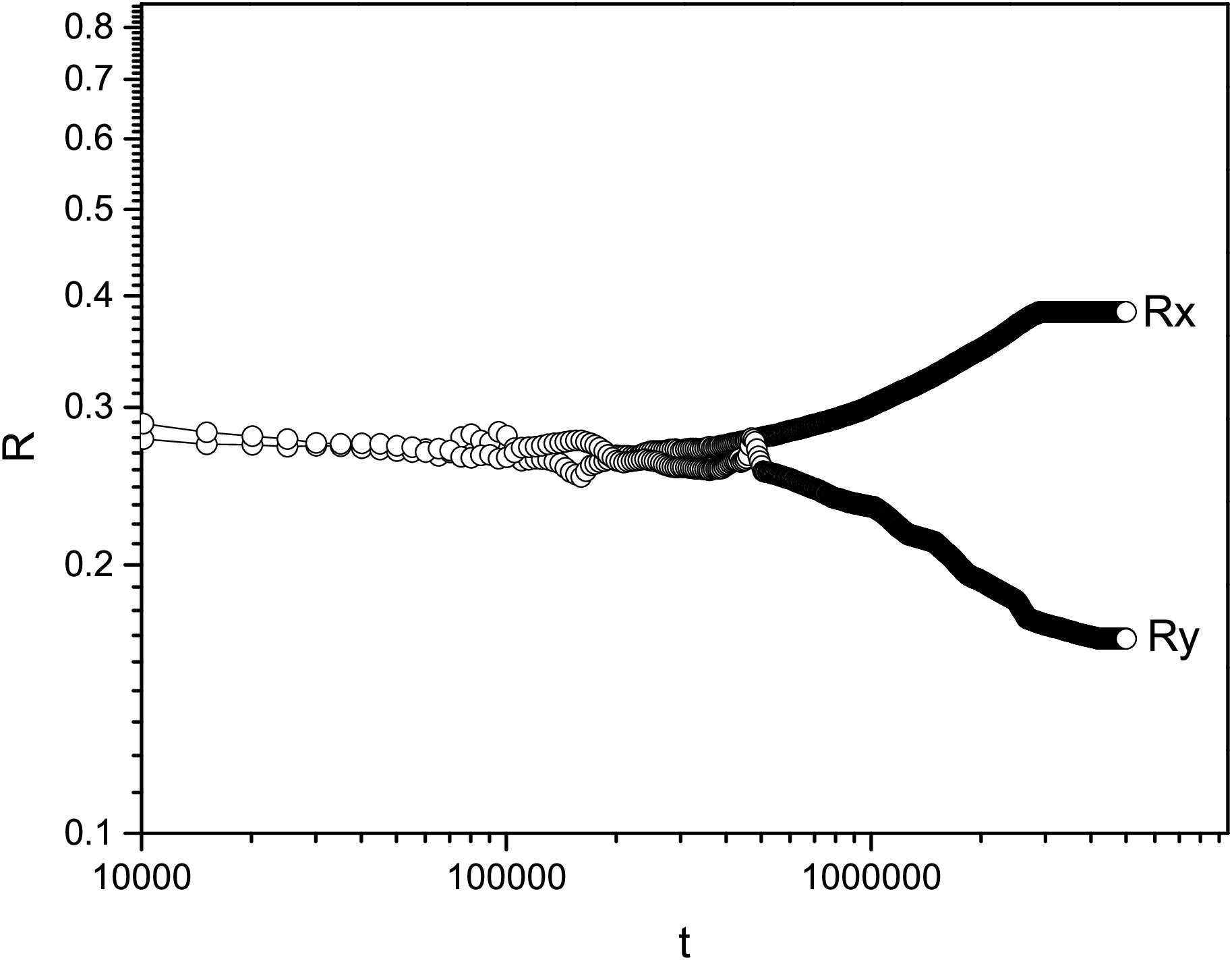}}
\caption{Growth of the domain size of the CD phase in the $x$ and $y$ directions as a function of time in double-logarithmic plots with $f_{AB/CD}=35/65$, $\omega=0.00001$, and $\gamma=0.032$.} \label{fig-smp7}
\end{figure}

In figure~\ref{fig-smp2}(a), with an increase in the shear frequency, the domain size $R_x$ of the AB microphase domain along the $x$ direction in the equilibrium state first increases (curves a--c) and then decreases (curves d--f). This indicates that the coarsening degree of the AB domain increases gradually along the oscillating field (first for small values of $\omega$) as the shear frequency increases. Then, when $\omega$ increases to a certain extent, the shear frequency is so high that the AB domains push each other, resulting in coarsening along the $y$ direction. At this moment, the coarsening along the direction of the oscillating shear field is inhibited. The coarsening perpendicular to the oscillatory shear of the AB phase becomes increasingly apparent, until it grows completely perpendicular to the oscillatory shear. By contrast, the domain size $R_y$ of the AB microphase domain along the $y$ direction first decreases (curves a--c) and then increases (curves d--f) as the shear frequency increases, as shown in figure~\ref{fig-smp2}(b). Furthermore, when the oscillating shear field increases to a certain extent, the AB phase occupying a larger proportion coarsens along the direction perpendicular to the oscillating shear field; however, this coarsening is inhibited when the oscillating shear field is weak. In addition, the obtained domain structure is stable, as indicated by the growth curve.

\subsection{Evolution progress}
We scrupulously examine the formation process of the lamellar structure with the AB phase (perpendicular lamellae) and CD phase (parallel lamellae) perpendicular to each other. Simulation snapshots obtained at different times for $\omega=0.66$ are shown in figure~\ref{fig-smp3}. The growth curves of the domain sizes along the $x$- and $y$-axes are presented in figure~\ref{fig-smp4}.

Figure~\ref{fig-smp3} presents the morphology evolution of the domain structure of the polymer system over time. In the evolution process, the macrophase separation is accompanied by microphase separation. The phase separation is not obvious at $t=250000$. At $t=500000$, there is an obvious microphase separation when the macrophase separation occurs. At this time, most of the AB phase is a tilted layered structure, which has a certain angle with the shear direction, except for a small part around the CD phase. The CD phase is disordered in the ring. At $t=2000000$, the AB phase tends to grow perpendicular to the oscillating shear field and becomes more ordered than before, while the CD phase is completely parallel to the direction of the oscillating shear field, forming a parallel layered structure. In the last stage, the AB phase grows completely perpendicular to the oscillating field and finally forms a layered structure of the AB phase (vertical layer) and the CD phase (parallel layer) perpendicular to each other.

Additionally, the evolution of the domain sizes in the $x$ and $y$ directions of the AB phase with the vertical layer and the CD phase with the parallel layer is examined, as shown in figure~\ref{fig-smp4}. Figure~\ref{fig-smp4}(a) presents the growth of the domain size of the AB phase as a function of time in a double-logarithmic plot, corresponding to the black and white regions in figure~\ref{fig-smp1}(f). Figure~\ref{fig-smp4}(b) presents the growth of the domain size of the CD phase as a function of time in a double-logarithmic plot, corresponding to the light gray and dark gray regions in figure~\ref{fig-smp1}(f). $R_y > R_x$ shown in figure~\ref{fig-smp4}(a) indicates that the domain size of the AB phase in the $y$ direction is considerably larger than that in the $x$ direction, which verifies the vertical layered structure of the AB phase in figure~\ref{fig-smp3}(d). Meanwhile, $R_x\gg R_y$ shown in figure~\ref{fig-smp4}(b) indicates that the domain size of the CD phase in the $x$ direction is much larger than that in the $y$ direction, which verifies the parallel layered structure of the CD phase in figure~\ref{fig-smp3}(d).

\subsection{Effect of oscillating shear field on composite system with different ratios}

To explore the effect of the concentration of DBC--DBC on the self-assembled structure, the initial concentration of the AB and CD block is changed to $f_{AB/CD}=35/65$. It is found that the self-assembly structure of the composite system under the oscillating shear field is completely different from that in $f_{AB/CD}=7/3$. Figure~\ref{fig-smp5} presents the microphase transformation of the composite system with an increase in the shear amplitude when the shear frequency is fixed at $\omega=0.00001$. As shown, the phase enclosed in the ring transforms from the original CD phase into the AB phase, and the macroscopic phase interfaces of both the AB and CD phases are the ring structure. When the shear amplitude is small, the AB phase forms the tilted layered structure far from the phase interface, while the CD phase is a bicontinuous structure [figures~\ref{fig-smp5}(a)--(b)]. The AB phase in the ring gradually becomes ordered along the direction of the oscillating shear field as the shear amplitude increases to 0.022 [figure~\ref{fig-smp5}(c)]. With a further increase in the oscillating shear field, the AB and CD phases of the composite system coarsen along the $x$ direction, forming an ordered parallel layered structure, except for the ring structure of the macrophase interface [figure~\ref{fig-smp5}(d)]. The parallel layered structure is disorganised, and the whole system is tilted in the $y$ direction [figures~\ref{fig-smp5}(e)--(f)].

The aforementioned results indicate that the influence of the oscillating shear field on the domain structure of the composite system is weak --- even negligible --- when the shear amplitude is small. The phase morphology of the composite system is essentially identical to that without the oscillating shear field. The oscillating field starts to take effect as the shear amplitude increases, and the phase separation of the system is disturbed. Then, the whole system tends to coarsen in the $x$ direction. When the shear amplitude increases to a certain extent, the system is completely parallel to the oscillating shear field except for the ring structure of the macrophase interface, exhibiting a perfect parallel layered structure. Finally, the amplitude is so large that the composite system accumulates in the direction perpendicular to the oscillating shear field, resulting in the whole system being tilted in the $y$ direction.

To further investigate the formation process of the parallel layered structure inside and outside the ring of the AB and CD phases, the evolution of the domain structure over time is discussed, and the growth curves of the CD phase in the $x$ and $y$ directions are analysed, as shown in figures~\ref{fig-smp6} and~\ref{fig-smp7}.

Figure~\ref{fig-smp6} presents the morphology evolution of DBC--DBC with $f_{AB/CD}=35/65$, $\omega=0.00001$, and $\gamma=0.032$. As shown, the phase separation of the system is not obvious, and macrophase separation first occurs at $t=10000$ [figure~\ref{fig-smp6}(a)]. Over time, microphase separation occurred simultaneously with macrophase separation. At this time, the AB phase is a lamellar structure inside the ring, while the CD phase is a bicontinuous layered structure [figures~\ref{fig-smp6}(b)--(c)]. At $t=3000000$, the system is anisotropic and exhibits a stable and ordered parallel layered structure, except for the ring structure of the macrophase interface.

Figure~\ref{fig-smp7} shows the growth of the domain size of the CD phase in the $x$ and $y$ directions as a function of time in double-logarithmic plots. $R_x\gg R_y$ indicates that the  domain size in the $x$ direction is far larger than that in the $y$ direction, corresponding to the structure in figure~\ref{fig-smp5}(d).

\section{Conclusions}

The self-assembly behaviour of DBC--DBC under an oscillating shear field is investigated via cell dynamics simulation. The results indicate that the macrophase separation of the composite system is accompanied by the corresponding microphase separation induced by the oscillating shear field. When the shear frequency increases, the AB phase changes from a tilted layered structure to a parallel layered structure, and finally to a vertical layered structure. The CD phase transforms from the initial concentric ring into a parallel layer in the ring and then into a parallel layered structure; thus, the system finally consists of a layered structure of the AB phase (vertical layer) and CD phase (parallel layer) perpendicular to each other. This structural transformation can be explained as follows. The increase in the shear frequency can accelerate the coarsening of the microdomain along the oscillating shear field; however, the different components push each other, resulting in coarsening perpendicular to the oscillating shear field when the frequency is too high. The AB phase occupying a larger proportion is significantly affected by the oscillating shear field, which results in an easy pileup in the direction perpendicular to the oscillating shear field, forming the perpendicular layered structure.
Since the CD phase is already confined in the ring, it breaks the restraint of the ring at a higher shear frequency and forms the parallel layered structure along the oscillating field. To verify this phase transition, the dynamic evolution of the domain size at different shear frequencies is analysed, revealing that it is identical to the previous phase transition. As for the formation of the layered structure of the AB phase (vertical layer) and CD phase (parallel layer) perpendicular to each other, the morphology evolution and the dynamic evolution of the domain growth are also examined. Moreover, the ordered phase transition with an increase in the oscillating shear field is different when the initial composition ratio of the system is changed. This conclusion provides a valuable guidance for the formation and transformation of ordered structures in experiments.

\section*{Acknowledgements}
Project supported by the Basic Research Plan of Shanxi Province, China (Grant No. 202103021223386), the Graduate Education Teaching Program of Shanxi Province, China (Grant No. 2022YJJG301).

\ukrainianpart

\title{Властивості самомонтажу системи диблок-кополімерів під дією осцилюючого зсувного поля}

\author{Й. Гуо\refaddr{label1}, Г. Хе\refaddr{label2}, Х. Фу\refaddr{label2}}

\addresses{
	\addr{label1} Факультет хімічної інженерії та матеріалознавства, Університет Люлянг, Ліші, 033001, Китай
	\addr{label2} Вища школа хімії та хімічної інженерії, Університет Шансі, Тайюань, 030006, Китай}

\makeukrtitle

\begin{abstract}
	\tolerance=3000%
Вивчаються властивості самомонтажу системи диблок-кополімерів під дією осцилюючого зсувного поля методом коміркового моделювання. Результати показують, що розділення макрофаз композитної системи супроводжується відповідним розділенням мікрофаз, викликаним осцилюючим зсувним полем. Зі збільшенням частоти фаза АВ змінюється від нахиленої шаруватої структури до паралельної і, насамкінець, до вертикальної. Фаза CD перетворюється з початкового концентричного кільця у паралельний шар у кільці, а потім --- у паралельну шарувату структуру. Таким чином, система остаточно формує шарувату структуру фази АВ (вертикальний шар) і фази CD (паралельний шар), які є взаємно перпендикулярними. Для перевірки фазового переходу аналізується зміна розміру домену при різних частотах зсувного поля. Упорядкований фазовий перехід зі збільшенням зсувного поля змінюється, коли міняється також початковий склад системи. Цей висновок може служити 
важливою настановою при формуванні та перетворенні впорядкованих структур в експериментальних умовах.
	\keywords диблок-кополімер, осцилююче зсувне поле, самомонтаж
	
\end{abstract}

\lastpage
\end{document}